\documentclass[10pt,conference]{IEEEtran}
\usepackage{cite}
\usepackage[dvips]{color}
\usepackage{tabu}
\usepackage{comment}
\usepackage{todonotes}
\usepackage{epsf}
\usepackage{epsfig}
\usepackage{times}
\usepackage{epsfig}
\usepackage{graphicx}

\usepackage{mathtools}
\usepackage{mathrsfs}
\usepackage{amssymb,amsmath}
\usepackage{amsmath}
\usepackage{amsfonts}

\usepackage{algorithmicx}  
\usepackage[algo2e]{algorithm2e} 
\usepackage{cleveref}
\usepackage{dsfont}
\usepackage{lettrine} 
\usepackage{epsfig,algorithm,algpseudocode,amsthm,url}
\usepackage{caption}
\usepackage{subcaption}
\usepackage[justification=raggedright]{caption}
\usepackage{subcaption}
\usepackage{bbm}
\usepackage[utf8]{inputenc}
\usepackage{verbatim}
\usepackage{setspace}	
\usepackage[T1]{fontenc}
\usepackage{textcomp}
\usepackage{tabularx}
\usepackage{xpatch}

\DeclareMathOperator*{\argmin}{argmin}
\usepackage{stackengine}
\usepackage[nocomma]{optidef}
\usepackage{commath}

\usepackage[left=1.62cm,right=1.62cm,top=1.85cm,bottom=2.6cm]{geometry}

\IEEEoverridecommandlockouts 

    \renewcommand{\baselinestretch}{0.9588}
\IEEEoverridecommandlockouts\IEEEpubid{\makebox[\columnwidth]{ 978-1-6654-5975-4/22/\$31.00~\copyright~2022 IEEE \hfill} \hspace{\columnsep}\makebox[\columnwidth]{ }} 

\begin{document}
\title{\Huge Evolutionary Deep Reinforcement Learning for Dynamic Slice Management in O-RAN
\thanks{
This material is based upon work supported by the Air Force Office of Scientific Research under award number FA9550-20-1-0090 and the National Science Foundation under Grant Numbers CNS-2232048.}
\thanks{\small F. Lotfi was with Department of Electrical and Computer Engineering, University of Colorado, Colorado Springs, CO, USA (e-mail: flotfi@uccs.edu), and now is with Department of Electrical and Computer Engineering, Clemson University, Clemson, SC, USA (flotfi@clemson.edu).

O. Semiari is with Department of Electrical and Computer Engineering, University of Colorado, Colorado Springs, CO, USA (osemiari@uccs.edu).

F. Afghah is with Department of Electrical and Computer Engineering, Clemson University, Clemson, SC, USA (fafghah@clemson.edu).} \vspace*{-0.4cm}}

\author{
\authorblockN{Fatemeh Lotfi, Omid Semiari, and Fatemeh Afghah}\\ \vspace*{-0em}
\vspace*{-0.9cm}}
\maketitle\vspace{-0.1cm}
\begin{abstract}
The next-generation wireless networks are required to satisfy a variety of services and criteria concurrently. To address upcoming strict criteria, a new open radio access network (O-RAN)  with distinguishing features such as flexible design, disaggregated virtual and programmable components, and intelligent closed-loop control was developed. O-RAN slicing is being investigated as a critical strategy for ensuring network quality of service (QoS) in the face of changing circumstances. However, distinct network slices must be dynamically controlled to avoid service level agreement (SLA) variation caused by rapid changes in the environment. Therefore, this paper introduces a novel framework able to manage the network slices through provisioned resources intelligently. Due to diverse heterogeneous environments, the intelligent machine learning approaches require sufficient exploration to handle the harshest situations in a wireless network and accelerate convergence. To solve this problem, a new solution is proposed based on evolutionary-based deep reinforcement learning (EDRL) to accelerate and optimize the slice management learning process in the radio access network's (RAN) intelligent controller (RIC) modules. To this end, the O-RAN slicing is represented as a Markov decision process (MDP) which is then solved optimally for resource allocation to meet service demand using the EDRL approach. In terms of reaching service demands, simulation results show that the proposed approach outperforms the DRL baseline by $62.2\%$.

\vspace{-0.cm}
\end{abstract}
\section{Introduction} \vspace{-0cm}
The next-generation wireless networks must be capable of managing a broad spectrum of wireless technologies with heterogeneous quality-of-service (QoS) and quality-of-experience (QoE) requirements. Among such technologies include emerging applications such as holographic telepresence, the Internet of everything, drone-based applications, and collaborative robots~\cite{6Gapplication,lotfi2022semantic,jebellat2021training,lotfi2021}. To meet strict QoS and QoE requirements of these applications, the new open radio access network (O-RAN) has been recently introduced which addresses the demand for virtual and programmable components, as well as intelligent, data-driven, and closed-loop control of the RAN. 

The key advantages of O-RAN are the possibility for operators to mix and match components, the use of open interfaces, the fact that it is software-centric and scalable, and the potential to improve network performance using machine learning (ML) approaches. Furthermore, all of these factors increase the flexibility of the network design~\cite{polese2022understanding}. The disaggregation of RAN functions into different units is a key feature of O-RAN that makes the network adaptable. The third generation partnership project (3GPP) splits base stations (BSs) into an open central unit (O-CU), an open distributed unit (O-DU), and an open radio unit (O-RU)~\cite{3gpp2017study}. Moreover, these distributed units with open interfaces link to RAN's intelligent controllers (RIC) to manage and control the network in near real-time and non-real-time controlling loops~\cite{oranslice2020,polese2022understanding}. The new generation networks need a controlling approach with fast convergence, between $10$ms and $1$s for near real-time controllers, to obtain their required QoS~\cite{oranAI}. Furthermore, to maintain the network QoS in the face of dynamic changes and heterogeneous requirements, O-RAN slicing is being explored as a viable solution. 

Different network slices need to be managed carefully to prevent service level agreement (SLA) diversity. While artificial intelligence (AI) and ML approaches are usable in network slicing, they face challenges, including the requirement for a vast amount of diverse data and sufficient exploration to accelerate convergence and train an ML model that can effectively generalize to different situations without impacting the actual RAN performance~\cite{polese2022understanding,brik2022deep}. Given the difficulty of gaining access to this amount and variety of data, making large-scale decisions involving several O-DUs will be challenging. The O-RAN slicing challenge to manage and control diverse and dynamic service requirements has recently been studied in several works~\cite{thaliath2022predictive,niknam2020intelligent,hammamipolicy,polese2021colo,bonati2021intelligence}. Two main ML categories that have been used in the literature are supervised deep learning and deep reinforcement learning (DRL). To avoid SLA violations, the works in~\cite{thaliath2022predictive} and~\cite{niknam2020intelligent} proposed supervised ML-based resource provisioning approaches for network slicing by using predictions on traffic and the number of active users in the network. The works in~\cite{hammamipolicy,polese2021colo,bonati2021intelligence} proposed DRL-based approaches to achieve online training and dynamic resource allocation in O-RAN. In particular, the authors in~\cite{hammamipolicy} provided a performance comparison between two on-policy and off-policy models for delay-tolerant and delay-sensitive services, and the work in~\cite{polese2021colo} utilized auto-encoders to minimize unnecessarily high-dimensional input for the DRL agent and improve DRL controller performance in the presence of unreliable real-time wireless data. Furthermore, the authors in~\cite{bonati2021intelligence} demonstrated the feasibility of RAN scheduling control over real-time RIC by collecting data at the network edge in a DRL method. 

Although the prior art in~\cite{thaliath2022predictive,niknam2020intelligent,hammamipolicy,polese2021colo,bonati2021intelligence} is effective in a set of use cases, it has a number of limitations including slow convergence and lack of generalizability. For example, the DRL methods in~\cite{hammamipolicy} require near 20k training time steps before they converge, and thus, they can lead to inefficient resource allocation in some delay-sensitive systems. In addition, while the work in~\cite{polese2021colo} aims to overcome the limitations of DRL approaches in real-time wireless network scenarios, it has not been completely resolved, especially in real-time controlling scenarios. 
While DRL algorithms are effective in complex tasks, their training procedures are slow in the face of unreliable real-time wireless data, causing delays in the O-RAN controlling mechanism. Furthermore, DRL methods suffer from a lack of sufficient exploration, particularly in dynamic heterogeneous environments such as the one studied in~\cite{bonati2021intelligence}. These challenges make DRL approaches insufficient for the O-RAN slicing scenario and mandate new solutions that can cope with the demand for broad exploration of dynamic wireless networks. 

The main contribution of this work is to utilize the opportunity provided by O-RAN to create new experiences based on disaggregated modules. 
To this end, we formulate a problem that aims to minimize the probability of SLA violation while considering some constraints on the total required resources. To solve this problem, we propose utilizing a joint DRL and population-based strategy as an evolutionary-based DRL (EDRL) approach in the O-RAN slicing scenario to accelerate the learning process in the RIC modules. By leveraging population actors, the EDRL provides enough exploration to stabilize convergence properties and make the learning process more robust. To do that, we consider the O-RAN slicing issue and model the O-RAN slicing managing problem as a Markov decision process (MDP). Then, to solve the MDP problem and find the best allocation policy for O-RAN slicing, we offer an EDRL algorithm. Simulation results show a $62.2$\% improvement in the network performance compared to the DRL baseline method. \emph{To the best of our knowledge, this is the first work that utilizes a hybrid method of evolutionary algorithms and DRL to achieve efficient and dynamic slice management in future O-RANs}. 

The rest of this paper is organized as follows. Section \ref{sysmodl} presents the system model and the problem formulation for the O-RAN slicing. Section \ref{EDRL} presents the proposed EDRL algorithm to solve the MDP problem. Simulation results are presented in Sec. \ref{sim} and conclusions are drawn in Sec. \ref{conclusion}.\vspace{-0.0cm}
\begin{figure}[t!]
  \centering
    \includegraphics[width=1\columnwidth]{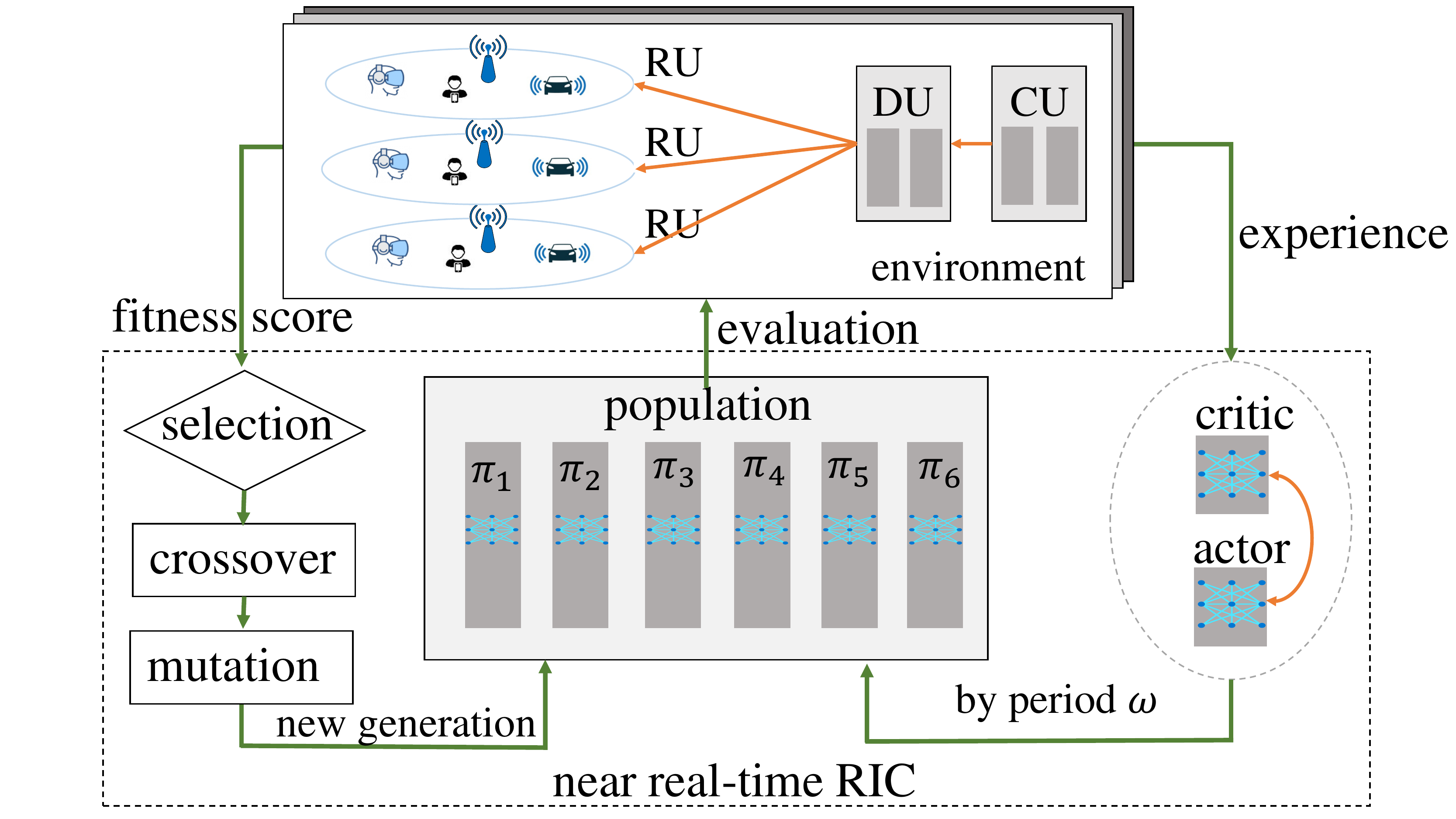}\vspace{-0cm}
    \caption{\small The proposed EDRL in Open RAN network.}\vspace{-0.2cm}
    \label{sys_graph}
\end{figure}
\section{System model}\label{sysmodl}
Consider an O-RAN architecture for a wireless network with $N$ heterogeneous users in a set $\mathcal{N}$ served by different network slices. By considering the O-RAN architecture as a dynamic slice optimization, we have resource management in RIC modules for O-RAN slicing as Fig \ref{sys_graph}. In this system, there are $L=3$ types of network slices in a set $\mathcal{L}$ with specific QoS requirements as $Q_l, l \in \mathcal{L}$, defined as follows: enhanced mobile broadband (eMBB),  machine-type communications (MTC), and ultra-reliable low latency communications (URLLC). Each of the slice $l \in \mathcal{L}$ serves $N_l$ user equipment (UE). Furthermore, the slices must share a pool of $K$ available resources to meet the QoS requirements of their assigned UEs. To guarantee that the SLA is satisfied, dynamic slice optimization will be performed in the RT-RIC module to optimize slice management. In addition, as a part of the slicing, the medium access control (MAC) layer should allocate resources following the radio resource management (RRM) strategy provided by the slice management. 
To address this challenge, next, we present the wireless model and, accordingly, formulate the slice management problem cognizant of wireless resource constraints.\vspace{-0cm}
\subsection{Wireless communication model for O-RAN network slicing}
According to the O-RAN architecture shown in Fig. \ref{sys_graph}, different network slices server different UEs with different QoS criteria. The RIC module is also in charge of managing these network slices and the resources assigned to them. Due to the stochastic nature of the wireless channel, static resource management would be ineffective. As a result, dynamic resource management is considered in order to dynamically re-assign resources to slice networks in each frame based on UEs' network and channel changes. As a result, while this aids in adapting to dynamic changes in the wireless channel, it also complicates resource assignment.

Besides, the QoS criteria $Q_l$ can also be defined particularly in terms of throughput, capacity, and latency for $\mathcal{L}$ slices.
To define the slice $l$'s achievable QoS, we consider orthogonal frequency-division multiple access (OFDMA) schemes. The slice $l$ achievable data rate can be written as\vspace{-0cm}
\begin{align}\label{urate}
    c_{l} = \lim_{\tau \to \infty} \frac{B}{\tau} \sum_{t=1}^{\tau}&\sum_{n=1}^{N_l}\sum_{k=1}^{K} e_{n,k} b_{l,k}\nonumber\\
    &\times\log\Big(1+\frac{p_u d_{n}(t)^{-\eta} h_{n,k}(t)}{ I_{n,k}(t)+ \sigma^2}\Big),
\end{align}
where $e_{n,k} \in \{0, 1\}$ is a binary variable that shows the resource block (RB) allocation indicator in the RB $k$ of user $n$, and $b_{l,k} \in \{0,1\}$ is a binary variable that shows the RB allocation indicator in the RB $k$ of slice $l$. Also, $B$ represents the RB bandwidth and $K$ is the total available RBs for downlink communications. In addition, $p_{u}$ is the O-RU transmit power per RB, and $d_{n}(t)$ is the user $n$ distance from its assigned O-RU. Moreover, $\eta$ represents the path loss exponent, and $h_{n,k}(t)$ is the time-varying Rayleigh fading channel gain. In \eqref{urate}, $I_{n,k}(t)$
denotes the downlink interference from the neighboring O-RUs transmitting over RB $k$, and $\sigma^2$ represents the variance of the additive white Gaussian noise (AWGN).
\subsection{Problem formulation}
Due to the restricted resources shared between slices with heterogeneous services and dynamic UEs, our goal is to minimize the probability of SLA violation $\mathbbm{P}(\abs{Q_l(\boldsymbol{e},\boldsymbol{b})-\lambda_l} \geq  \epsilon_l)$, $\forall l \in \mathcal{L}$, while considering some constraints on the total required resources.  
Also, $\boldsymbol{e}$ and $\boldsymbol{b}$ represent vectors of $e_{n,k}$ and $b_{l,k}$ as resource allocation indicators at MAC and RIC, respectively. 
Therefore, we formulate the following optimization problem to find an optimal allocation policy for O-DUs and distribute the shared resources among $N_l$ heterogeneous UEs in each O-RAN slice $l$.
\begin{subequations} 
\begin{align}\label{opt1}
 \argmin_{\boldsymbol{b},\boldsymbol{e}} & \hspace{0.5cm} 
 \mathbbm{P}(\abs{Q_l(\boldsymbol{e},\boldsymbol{b})-\lambda_l} \geq  \epsilon_l),\\
 \text{s.t.,} 
& \hspace{0.5cm} \forall l \in \mathcal{L},\,\,  \forall n \in \mathcal{N} , \label{opt1_q}\\
& \hspace{0.5cm}  \sum_{l=1}^{L}\sum_{n=1}^{N_l}\sum_{k=1}^{K} b_{l,k}e_{n,k} \leq K, \label{opt1_Ns}\\
& \hspace{0.5cm} \sum_{l}b_{l,k} \leq 1,\,\,   \label{opt1_e}
\end{align}
\end{subequations}
where $\lambda_l$ and $\epsilon_l$ represent, respectively, the desired threshold and margin values in QoS required by slice $l$. Constraint \eqref{opt1_Ns} and \eqref{opt1_e} represent the feasibility conditions on allocated resources to slices and UEs.  
Problem \eqref{opt1} indicates that the $\mathcal{L}$ slices achieve their demanded QoS by minimizing the probability of SLA violation. The proposed problem is an NP-hard mixed-integer stochastic optimization problem which is challenging to solve. 
Markov decision process (MDP) provides a mathematical framework for decision-making and optimization problems involving partially random situations. As a result, it is advantageous to model the \eqref{opt1} as an MDP and solve it using dynamic methods such as DRL approaches.\vspace{-0.0cm}


\subsection{Stochastic game-based optimization problem}
By considering the RT-RIC module as an intelligent agent that makes decisions to manage the O-RAN slicing, the other components, i.e., O-RU, O-DU, and O-CU will act as the agent's environment that is influenced by the agent's actions, as shown in Fig \ref{sys_graph}. Thus, the decision process of the mentioned O-RAN slicing controlling the problem is represented as an MDP with tuples $\langle \mathcal{S},\mathcal{A}, T,\gamma,r \rangle$, where $\mathcal{S}$, $\mathcal{A}$, and $T$ represent the state space, action space, and transition probability from current state to the next state, respectively. The MDP tuples are described as follows:
\subsubsection{State} $s_t \in \mathcal{S}$ represents the O-RAN status in each step of time which contains the achievable QoS of each slice $Q_l$, UEs' density in each slice $N_l(t)$, and resource allocation history as the previous action $a_{t-1}$. Therefore, the observation of the intelligent agent in time $t$ is as follows:  
    \begin{equation}
        s_t = \{Q_l,N_l,a_{t-1}\mid \forall l\in \mathcal{L}\}.
    \end{equation}
\subsubsection{Action} $a_t \in \mathcal{A}$ is defined as a vector of the number of required resources for the O-RAN slices and UEs. Thus, in each time $t$, the agent, based on its policy, decides to perform the action as $a_t = \{\boldsymbol{e},\boldsymbol{b}\}$.
\subsubsection{Reward} $r_t$ characterizes summation of the complement probability of SLA violation in \eqref{opt1}, which relies on the incoming traffic of each slice and the radio condition of the connected UEs. The desired reward value is described as:
   \begin{equation}
        r_t = \sum_{l=1}^{\abs{\mathcal{L}}}\Big(1- \mathbbm{P}(\abs{Q_l(\boldsymbol{e},\boldsymbol{b})-\lambda_l} \geq  \epsilon_l) \Big).
    \end{equation}
Therefore, the procedure makes best use of the available bandwidth to meet the QoS demands of all slices. Then, the defined MDP model can be investigated using a DRL approach. The main task of a DRL agent is to find an optimal policy $\pi^*(a_t|s_t;\theta_p)$ as a mapping from the state space to the action space that maximizes its expected average discounted reward $\mathbb{E}_{\pi}[R(t)]$, where $R(t) = \sum_{i=0}^{\infty}\gamma^i r_{i,t}$. Given a policy $\pi$, the state-value and action-value functions are defined as $V(s_{t}) = \mathbb{E}_{\pi}[R(t)\mid s_{t}]$ and $Q(s_{t},a_{t}) = \mathbb{E}_{\pi}[R(t)\mid s_{t},a_{t}]$, respectively. Due to continuous states and actions, the Deep Deterministic Policy Gradient (DDPG) has been used as a model-free and off-policy algorithm. As part of the actor-critic technique, the RT-RIC agent simultaneously learns an ideal policy to assign the resources that optimize long-term reward. The policy network parameterized by $\boldsymbol{\theta}_p$ is updated using the gradient defined as follows with $e$ random samples transitions:
\begin{align}\label{pupdate}
    \nabla_{\theta_p}J = \frac{1}{e}\sum_{i=1}^{e} \nabla_a Q\big(s_i,a_i;\theta_v\big) \nabla_{\theta_p}\pi\big(a_i,s_i;\theta_p\big).
\end{align}
Also, the value network parameterized by $\boldsymbol{\theta}_v$ will be updated by minimizing the following loss as:\vspace{-0.1cm}
\begin{align}\label{vupdate}
    \min_{\theta_v} \frac{1}{e}\sum_{i=1}^{e} \bigg(r_i + \gamma Q_{\pi_{i+1}}(s_{i+1},a_{i+1};\theta_v)-Q_{\pi_i}(s_{i},a_{i};\theta_v)\bigg)^2. \vspace{-0.1cm}
\end{align}
Training one agent that interacts with the environment can be very time-consuming. However, with O-RAN, we may leverage experience from all of the disaggregated modules to guide the agent in the training procedure. For instance, different O-DUs may experience similar network instances (i.e., network traffic, QoS requirements, etc.) while being deployed at different locations across the network. 
Accordingly, sharing their experiences increase the generality of the resource assignment task. The prior works in \cite{thaliath2022predictive,hammamipolicy,polese2021colo,niknam2020intelligent,bonati2021intelligence} utilize the supervised deep learning and DRL approaches for network management. 
The method used in these works has limitations such as the need for large-scale data, lack of enough exploration, and unstable convergence of the supervised deep learning and DRL approaches in O-RAN slicing. Hence, inspired by \cite{khadka2018evolution} we employ a hybrid strategy to solve the problem \eqref{opt1}. This hybrid strategy combines the evolutionary algorithm (EA) optimization method with the DDPG algorithm of the DRL approach to better utilize experience samples and provide a more effective performance in less time than DRL alone. 
EA provides a diverse set of samples representing a wide range of services and traffic requirements to improve DRL learning performance. In response, DRL injects gradient information into the EA population. Injecting DRL gradient information into EA, augments EA's ability to select samples that force the DRL in policy space toward the regions with higher reward.\vspace{-0.1cm}
\section{Evolutionary DRL method}\label{EDRL}
In fact, EDRL is a hybrid method combining population-based EA and high sample efficiency DRL approaches. The EDRL uses diverse EA experiences to train the DRL, while DRL injects gradient information into the EA population. Accordingly, it makes them powerful to converge faster and thus are suitable for real-time applications~\cite{oranAI}. The employment of a fitness metric that aggregates returns throughout a whole episode makes EAs beneficial in environments with reward-less states where the reward is only specified and known for a few states and is resistant to long time horizons~\cite{khadka2018evolution}. Accordingly, the EDRL addresses the delayed reward issue, which is obvious in network slicing since the network requires to experience diverse policies to offer a different reward. In general, the flow of the EDRL algorithm is divided into three interacting phases; the population phase, the DRL phase, and the interaction between them which are explained in the following paragraphs. 

\subsection{Population phase} The population actors $\pi_{\textit{EA},i}$ \emph{evaluate} in one episode of interaction with the environment during the population phase, as illustrated in Fig. \ref{sys_graph}. During the evaluation episode, they measure the fitness score as a cumulative sum of return value in each interaction $t$ as $F_i = \sum_{t}r_{i,t}$, and save the actors experience in replay buffer $\mathcal{B}$ as tuple $\langle s_t,a_t,s_{t+1},r_t \rangle$. As is clear, the measured fitness score is determined by achieved QoS of each slice as $Q_l$ which depends on the wireless aspects of the environments. Then, based on the value of the fitness scores, the population actors get sorted for \emph{selection} part. Consequently, the results will be used in the \emph{mutation} and \emph{crossover} sections to create the next generation using the elite individuals of the population. Here, the O-RAN allows providing new experiences through disaggregated modules in different geographical locations, such as leveraging different populations with separate environments. As a result, the network will experience numerous wireless communication traffics, which is crucial to improve the generalization ability of the network's dynamic management system.  

\subsection{DRL phase} On the other hand, in the DRL phase, a critic network that is parameterized by $\theta_{v}$ will be updated using a random batch of replay buffer $\mathcal{B}$ samples by gradient descent (GD) manner as \eqref{vupdate}. Then, the critic network trains a DRL actor $\pi_{\text{RL}}$ by sampled policy gradient \eqref{pupdate}. The EA simply uses samples in the fitness score and then leaves information. However, by utilizing them in a replay buffer and continually applying powerful gradient-based actor-critic algorithms, they extract more information from data while maintaining high sample efficiency. 

\subsection{Interaction between populations and DRL } The most crucial phase will then be performed, which include an interaction between the EA and DRL algorithms. During this phase, the $\pi_{\text{RL}}$ weights are copied to the worst-performing individuals in the population actors $\pi_{\text{EA}, i}$ and cause leverage of the learned information from DRL and help to stabilize learning. Thus, they learned from the experience of episodes as well as fitness scores by taking this approach. Also, a synchronization period $\omega$ governs how frequently the RL-actors information shares with the EA population. Furthermore, following the selection of the elites, the $\pi_{\text{RL}}$ is updated by the best performers of $\pi_{\text{EA}, i}$ to accelerate convergence. 

While the population actors explore in parameter space, and the RL actor explores in action space, they complement each other and lead to effective policy space exploration. 
Besides, the critic network is updated with samples from the EA population's policy, which may or may not be used in the DRL agent's next action. As a result, this hybrid method behaves as an on-policy method in each synchronization period, $\omega$, and as an off-policy method other times, providing benefits from both methods~\cite{gu2017interpolated}. 
The effectiveness of EA is determined by the selection, the crossover, and the mutation parts. It means that the appropriate choice of these parts is critical. The parents will be chosen in the selecting part to generate the next generation. As is obvious, selecting for higher fitness scores will improve overall quality in each generation. On the other hand, selection by diversity avoids to stuck in a local extremum. The selection would be determined by taking into account the environment. In our system model, the tournament selection~\cite{miller1995genetic} would have the optimum performance while considering the O-RAN environment as follows:
\begin{equation}\label{select}
    \pi_{\text{EA},p1},\pi_{\text{EA},p2} =\text{tournament}(\pi_{\text{EA}}),
\end{equation}
where $p1$ and $p2$ represents the selected parents to generate the next generation. In the O-RAN environment, the populations are actor-networks that are located in the RT-RIC modules and have interaction with distinct O-DU/O-RU environments in different areas as Fig. \ref{sys_graph}. 
Following parent selection, the crossover and mutation sections inject additional randomness into the system, resulting in more exploration and generalization. To ensure the transfer of parents' valuable properties to the next generation, we use the average function in the crossover section as follows: 
\begin{equation}\label{cross}
    \hat{\theta}_{\text{EA}} = \text{avg}(\theta_{\text{EA},p1},\theta_{\text{EA},p2}).
\end{equation}
Then in the mutation part, the population would be perturbed to create new features for the next generation as follow~\cite{khadka2018evolution}:
\begin{align}\label{mut}
    \Tilde{\theta}_{\text{EA}} = \begin{cases}
     \hat{\theta}_{\text{EA}} + \mathcal{N}(0,100 \xi), & \kappa \leq q_{\text{super-mut}},\\
     \mathcal{N}(0,1), & \kappa \leq q_{\text{reset}},\\
      \hat{\theta}_{\text{EA}} + \mathcal{N}(0,\xi), & \text{otherwise},    
    \end{cases}
\end{align}
where $\mathcal{N}(0,1)$ represents a Gaussian noise with zero mean and unit variance. Also, $\xi$ shows the mutation rate, and $\kappa$ is a random variable in $[0,1]$. Moreover, the $q_{\text{super-mut}}$ and $q_{\text{reset}}$ are super mutation probability and reset probability, respectively. 
\begin{algorithm}[t!]
\SetAlgoLined
\textbf{Input}: $N_g$,\,\,$N_p$,\,\,$N_e$,\,\,$\theta_{\text{EA},i},\forall i \in [0,N_p]$,\,\,$\kappa$,\,\,$\omega$,\,\,$\theta_{p}$,\,\,$\theta_v$.    \\
\For{iteration $g=1:N_g$}{
\For{actor $i=1:N_p$}{
$F_i = \text{evaluate}(\pi_{\text{EA},i})$.\\
$\mathcal{B}\gets \langle s_t,a_t,s_{t+1},r_t \rangle $.
}
elite individuals $\mathcal{E} =\{\pi_{\text{EA},i}\mid \max_i(F_i,N_e)\}$.\\
$\text{Evolution}\big(\mathcal{E},\eqref{select},\eqref{cross},\eqref{mut}\big)$.\\
Update $\theta_v$ using \eqref{vupdate} and a random batch of $\mathcal{B}$.\\
Update $\theta_p$ using \eqref{pupdate}.\\
\If{$g \mod{\omega} =0$}{
 Copy the DRL actor weight $\theta_p$ into the weakest individual in population $\theta_{\text{EA},i} \gets \theta_p$.
 }
 \If{$\theta_p$ by \eqref{pupdate} is converged}{
 Break.
 }
}
\textbf{Output}: $\theta_p$,\,\,$\theta_v$. \\
\caption{The EDRL algorithm}\vspace{-0cm}
\label{alg1}
\end{algorithm}\vspace{-0.2cm}
\subsection{The proposed EDRL algorithm}
In Algorithm \ref{alg1}, we summarize the EDRL method to solve the optimization problem in \eqref{opt1}-\eqref{opt1_e}. The input variables of the algorithm are the number of generation $N_g$, the population size $N_p$, the number of elites $N_e$, and the population actors' network weights as $\theta_{\text{EA},i}$ which are initialized with random weights. Also, the DRL agent's actor and critic network weights as $\theta_p$ and $\theta_v$ are initialized randomly. Moreover, a random variable $0\leq \kappa \leq 1$ and a synchronization period $\omega$ are considered as input variables in Algorithm \ref{alg1}. The algorithm proceeds to output the trained policy of the DRL agent. In each generation loop $g$, the population actors' fitness score $F_i$ is measured, and each population actor experience is stored in the replay buffer $\mathcal{B}$. Then, in the "Evolution" section of the Algorithm, the $N_e$ of the best population actors based on the measured $F_i$ are selected as elites $\mathcal{E}$ to generate the next generation of population actors based on the parent selection \eqref{select}, crossover \eqref{cross} and mutation \eqref{mut}. In this step, the DRL agent updates its actor and critic networks by using \eqref{pupdate}, \eqref{vupdate}, and the replay buffer random samples. The algorithm terminates once the DRL agent policy network is converged or after $N_g$ maximum generation. \vspace{-0cm}

\begin{table}[t!] 
	\footnotesize
	\centering
	\caption{\vspace*{-0cm} Simulation parameters }\vspace{-0.0cm}
	\begin{tabular}{|>{\centering\arraybackslash}m{2cm}|>{\centering\arraybackslash}m{4cm}|}
		\hline
		\bf{Parameter} &\bf{Value } \\
		\hline
		Subcarrier spacing & $15$ kHz\\
		\hline
		Total bandwidth & $10$ MHz  \\
		\hline
		RB bandwidth  & $200$ kHz\\
		\hline
		$p_u$ & $56$ dBm \\
		\hline
		$K$ & 50 \\
		\hline
		$h$ & 10-tap Rayleigh fading channel \\
		\hline
		$\sigma^2$ & $-173$ dBm \\
		\hline
		DDPG batch size & 128  \\
		\hline
		Replay buffer size & $1e6$  \\
		\hline
		$q_{\text{super-mut}}$ & $0.05$ \\
		\hline
		$q_{\text{reset}}$ & $0.1$ \\
		\hline
		Mutation probability & $0.9$ \\
		\hline
		 $\omega$ & $10$ \\
		\hline
		$N_g$ & $100$ \\
		\hline
		$N_p$ & $10$ \\
		\hline
	\end{tabular}\label{param} \vspace{-0cm}
\end{table}

\section{simulation results}\label{sim}
In the simulation part, we investigate an O-RAN architecture with three slices (i.e., eMBB, MTC, URLLC). We consider a $10$ MHz bandwidth as $50$ RBs which can be dynamically assigned to different slices. The slices serve a total of $30$ users which are randomly and uniformly distributed in the network and divided among the slices as $5$, $20$, and $5$. In the EDRL part, we consider the $10$ actor as the population, with elite fraction $0.2$ and crossover and mutation batch size $128$ and $256$, respectively. In the DRL part, we implement the DDPG algorithm using Pytorch with three fully-connected layers with $128$, $256$, $256$ neurons for actor and critic networks and a \textit{tanh} activation function. For all the models, the learning rate $10^{-4}$ and \textit{Adam} optimizer are considered. Table \ref{param} summarizes the simulation parameters. 

\begin{figure}[t!]
  \centering
    \includegraphics[width=7.9cm]{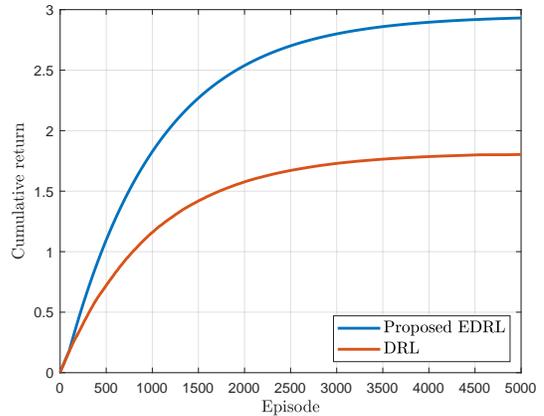}\vspace{-0.0cm}
    \caption{\small Performance comparison of EDRL and DRL algorithm.}\vspace{-0.2cm}
    \label{cum_return}
\end{figure}

Figure \ref{cum_return} compares the performance of the proposed EDRL algorithm with the DRL algorithm as a baseline method. The cumulative rewards in Fig. \ref{cum_return} are measured with $\gamma=0.95$. The results are displayed in each episode, for comparison with conventional DRL and are averaged over a sufficient number of runs. The results in Fig. \ref{cum_return} reveal that the EDRL approach can give up to a $62.2\%$ greater final return value than the DRL method and prove the efficacy of the proposed EDRL system over the wireless environment. Fig. \ref{cum_return} also illustrates that the suggested approach achieves faster convergence than the baselines, that is due to the employment of a population-based method with $10$ actors, who supply many valuable experiences of diverse modules to use in the training of RL-based actors. Furthermore, as Fig. \ref{cum_return} indicates, the DRL method which has better sample efficiency shows better performance at the early episodes of the training process, where limited samples are available. After some episodes, the joint DRL and population-based method EDRL, which has experienced more samples in comparison to early episodes, provide more generality in the training process and thus outperforms the DRL.

\begin{figure}
     \centering
     \begin{subfigure}[a]{0.18\textheight}
         \centering
         \includegraphics[width=\textwidth]{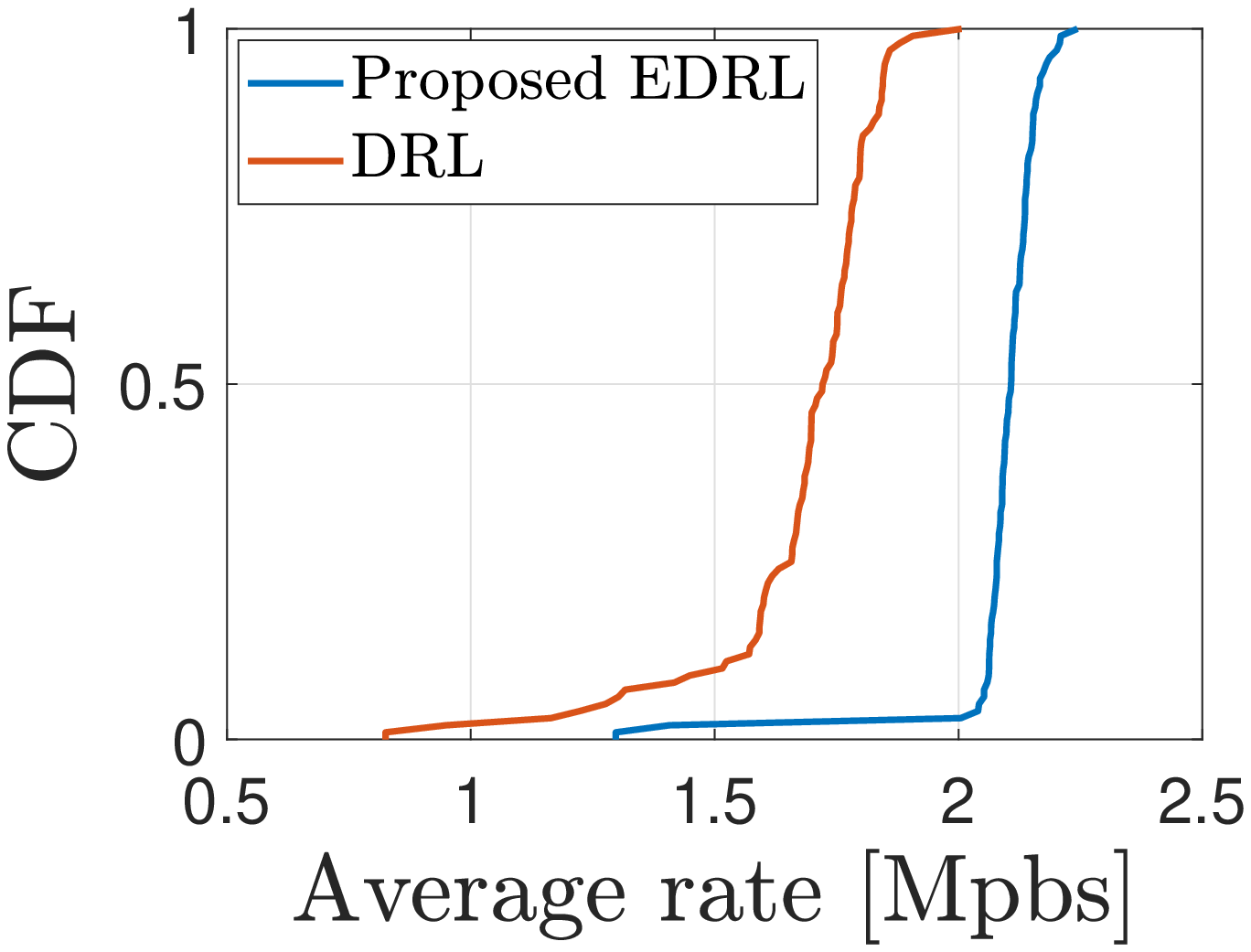}
         \caption{Qos in slice 1}
         \label{cdf_q1}
     \end{subfigure}
     \hfill
     \begin{subfigure}[a]{0.18\textheight}
         \centering
         \includegraphics[width=\textwidth]{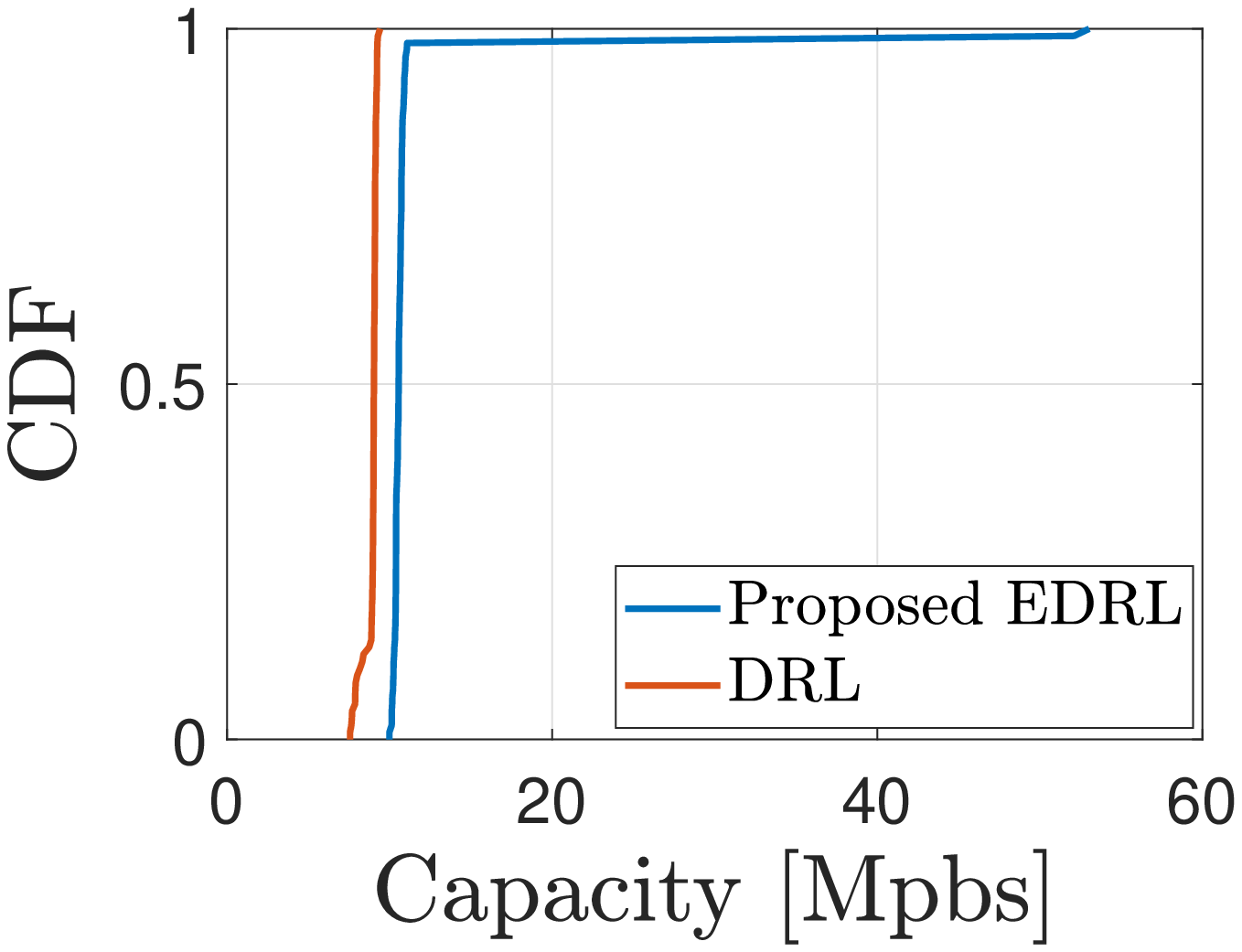}
         \caption{Qos in slice 2}
         \label{cdf_q2}
     \end{subfigure}
     \hfill
     \begin{subfigure}[a]{0.18\textheight}
         \centering
         \includegraphics[width=\textwidth]{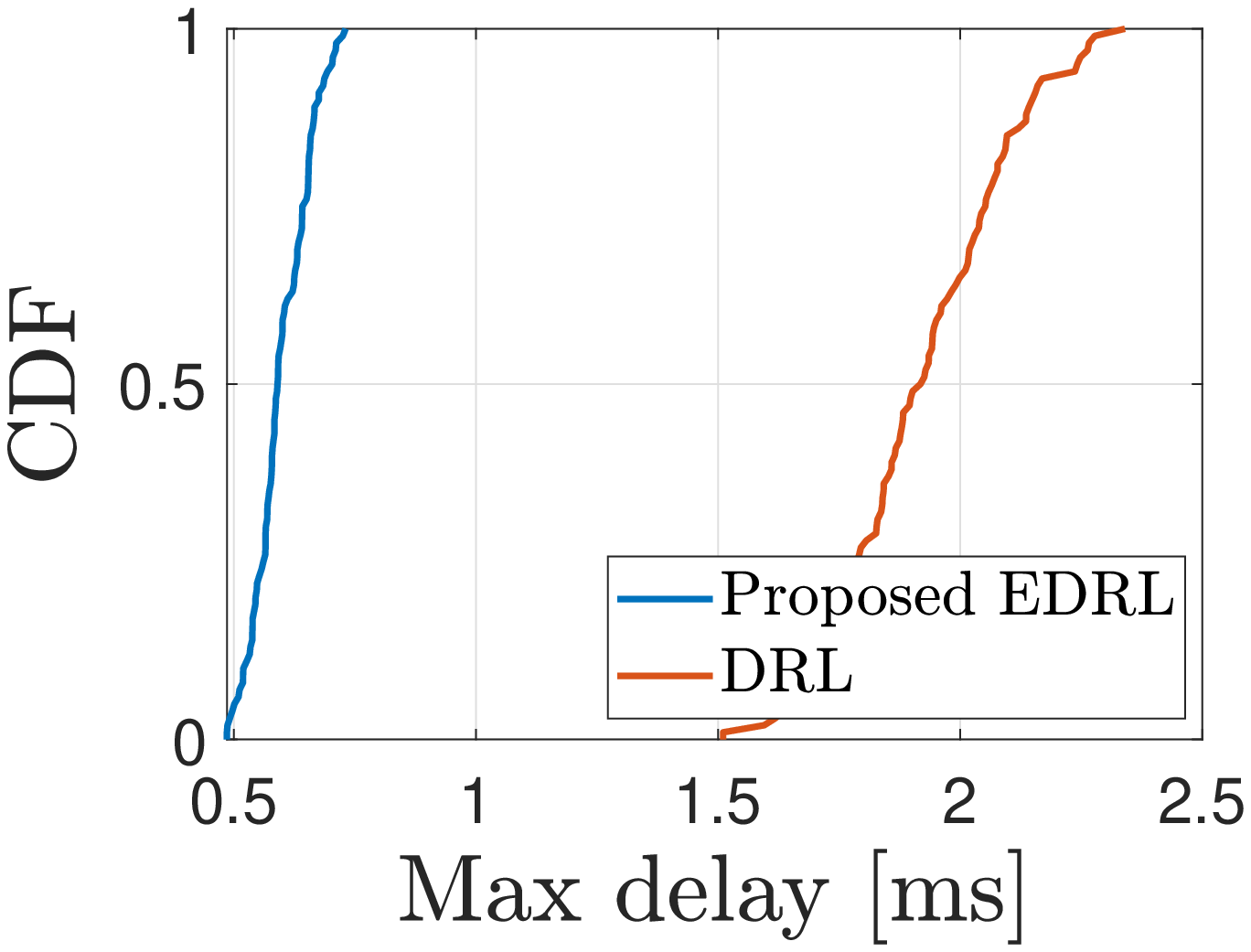}
         \caption{Qos in slice 3}
         \label{cdf_q3}
     \end{subfigure}
        \caption{CDF of achieved QoS in each slice through EDRL and DRL training process.}\vspace{-0.2cm}
        \label{cdf_qos}
\end{figure}

Figure \ref{cdf_qos} shows the Cumulative Distribution Functions (CDFs) of the achieved QoS in each slice through the EDRL and DRL training process. The slices are assumed to be the eMBB slice, MTC slice, and URLLC slice, in that order. Each slice meets the service demands by considering a distinct QoS for specific services, such as the average data rate in the eMBB slice in Fig. \ref{cdf_q1}, the capacity of the MTC slice in Fig. \ref{cdf_q2}, and the maximum delay in URLLC slice in Fig. \ref{cdf_q3}. Considering an average data rate QoS metric in eMBB slices guarantees a stable service for connected UEs. Similarly, the capacity as a QoS parameter in MTC slices ensures that a high connection density is supported. Furthermore, considering an exponential random variable with an average size of $10$ Kb as packet length for URLLC slice and selecting maximum delay as a QoS criterion ensures the lowest possible value for maximum latency in these delay-sensitive services. 

\begin{figure}
     \centering
     \begin{subfigure}[a]{0.18\textheight}
         \centering
         \includegraphics[width=\textwidth]{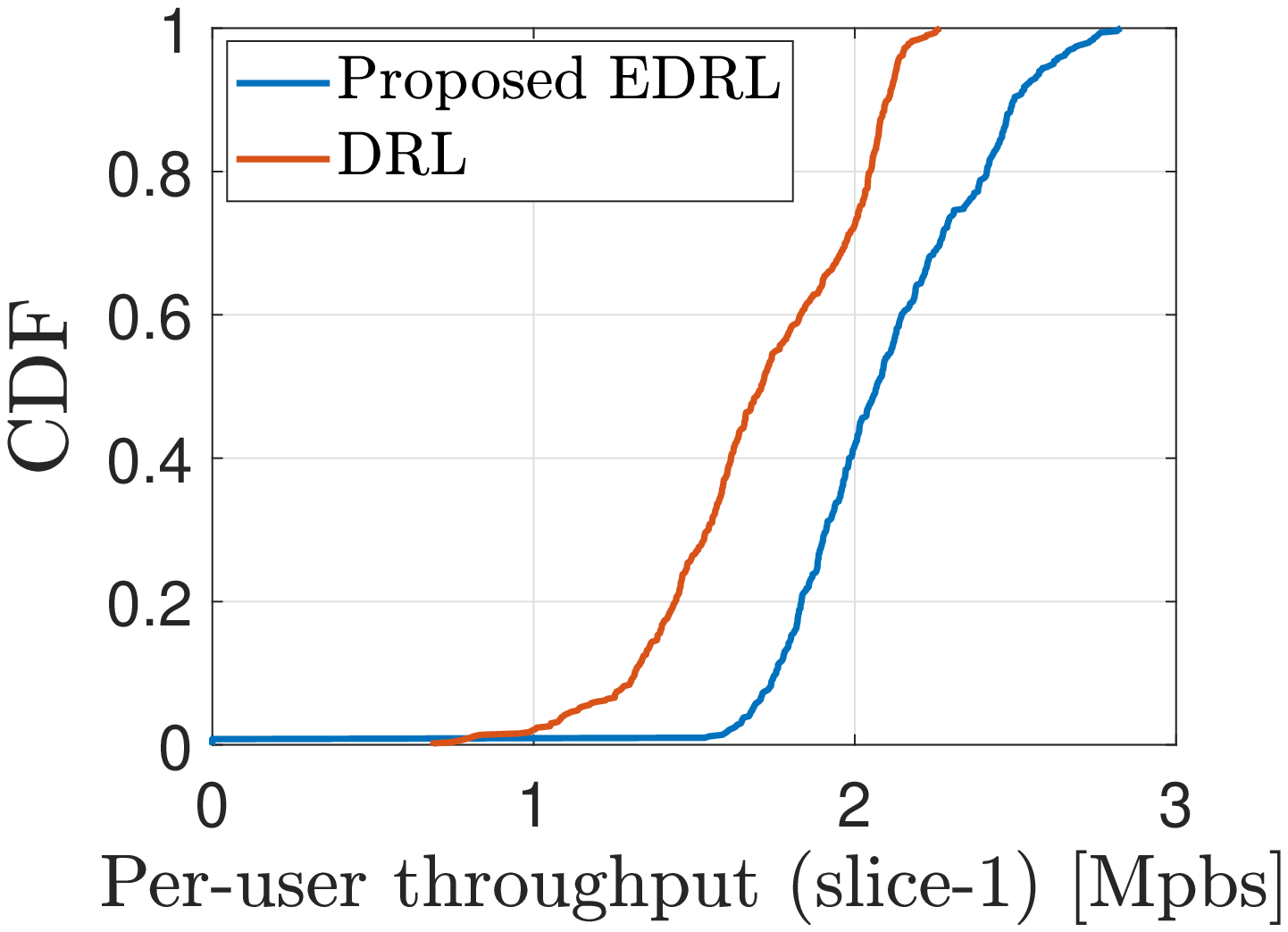}
         \caption{Slice 1 UEs}
         \label{cdf_ue1}
     \end{subfigure}
     \hfill
     \begin{subfigure}[a]{0.18\textheight}
         \centering
         \includegraphics[width=\textwidth]{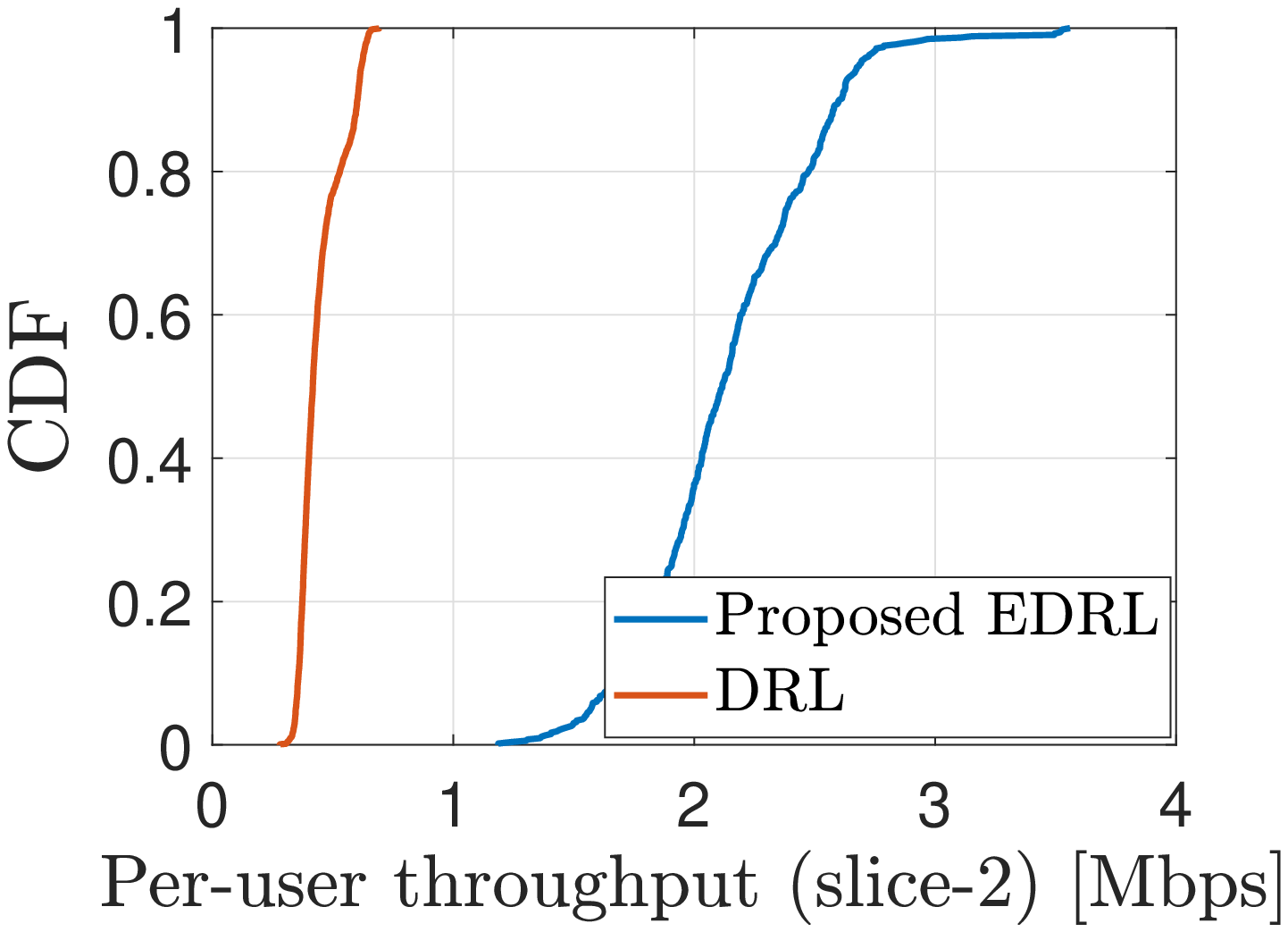}
         \caption{Slice 2 UEs}
         \label{cdf_ue2}
     \end{subfigure}
     \hfill
     \begin{subfigure}[a]{0.18\textheight}
         \centering
         \includegraphics[width=\textwidth]{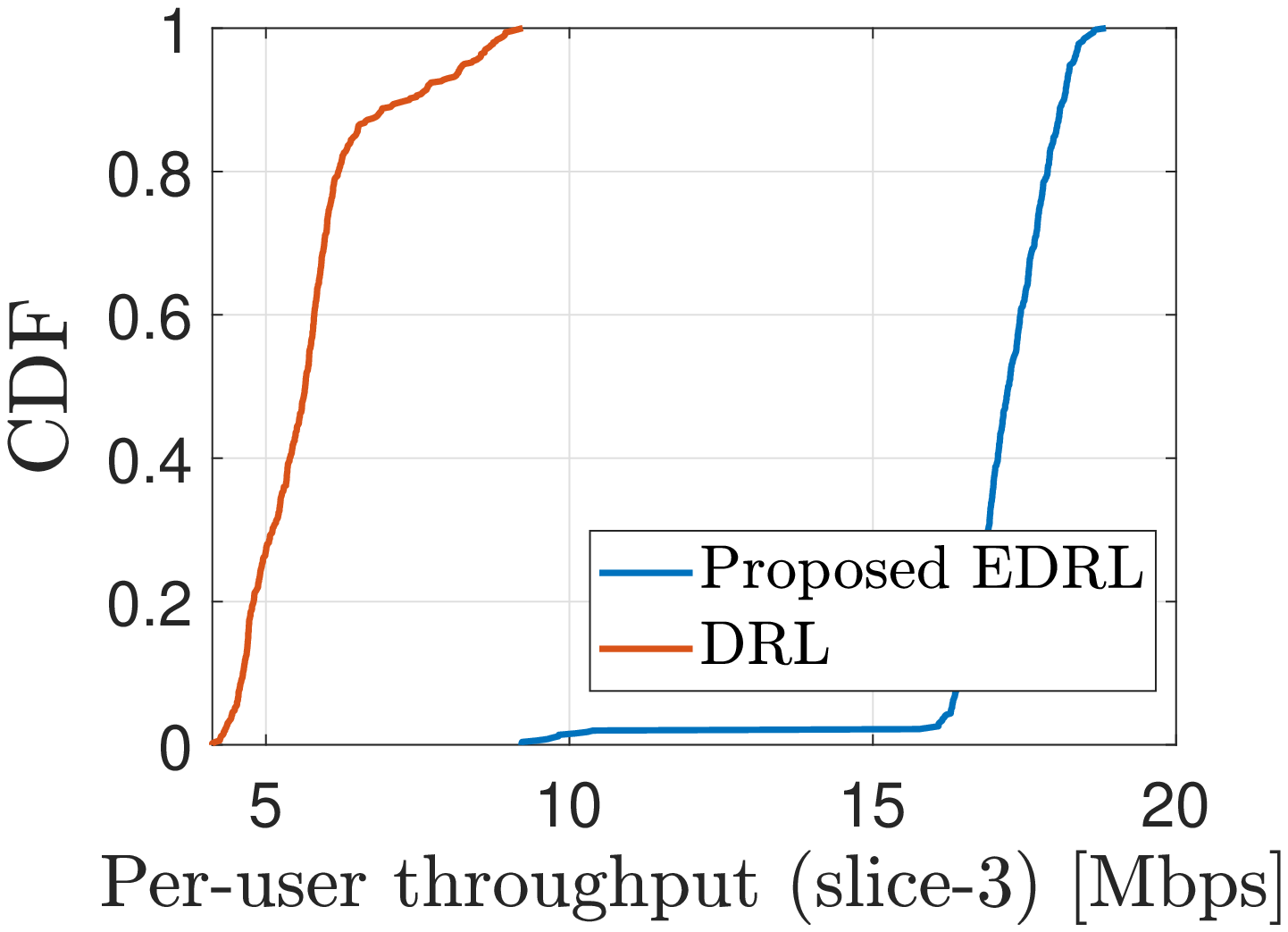}
         \caption{Slice 3 UEs}
         \label{cdf_ue3}
     \end{subfigure}
        \caption{CDF of achieved throughput per users through EDRL and DRL training process.}\vspace{-0.2cm}
        \label{cdf_UE}
\end{figure}

Figure \ref{cdf_UE} shows the per-user throughput in each slice of the simulation O-RAN environment. The results presented in Fig. \ref{cdf_UE} were obtained during $100$ generation iterations by distributing $5$, $20$, and $5$ users to the slices, respectively. As seen in Fig. \ref{cdf_UE}, the network users follows the QoS demand in Fig. \ref{cdf_qos}. Furthermore, the results indicate that the Algorithm \ref{alg1} was successful in keeping the users' throughput within the intended range. \vspace{-0.1cm}

\section{Conclusion}\label{conclusion}
In this paper, we have developed a novel O-RAN slicing framework over an evolutionary-based DRL approach. To this end, we have formulated an optimization problem that efficiently allocates the shared available RBs to each O-RAN slice. To solve this problem, we have modeled the optimization problem as an MDP, then by employing the desegregated modules in O-RAN architecture, we develop a new EDRL algorithm to find an optimal policy for allocating the available resources to distinct O-RAN slices. Accordingly, utilizing the population experiences in the DRL training, the trained policy is more general and robust in different traffic situations over the wireless networks. The simulation results have shown up to $62.2\%$ improvements in maximum rewards compared to the DRL baseline method. Further, the results have highlighted the importance of utilizing different experiences and generalization in policy training in dynamic wireless networks, and demonstrated the efficiency of the proposed algorithm in presence of wireless bandwidth constraints.
 \vspace{-0cm}

\def\baselinestretch{0.96}
\bibliographystyle{IEEEbib}
\bibliography{Main}
\end{document}